\documentclass[a4paper,11pt]{article}
\pdfoutput=1 
\usepackage{jheppub} 
\usepackage[utf8]{inputenc}
\usepackage{amsthm}
\usepackage{mathrsfs} 
\usepackage[all]{xy}
\graphicspath{{./images/}}
\theoremstyle{plain}

\theoremstyle{definition}

\numberwithin{thm}{section}

\def\d{{\rm d}}
\def\i{{\mathsf i}}

\usepackage{slashed}

\def\cZ{{\cal Z}}

\def\bR{{\mathbb R}}

\def\bZ{{\mathbb Z}}

\def\sc{{\mathsf c}}

\def\sq{{\mathsf q}}

\def\U{\mathrm{U}}

\def\SO{\mathrm{SO}}

\def\beq#1\eeq{\begin{align}#1\end{align}}

\def\w{\widetilde}

\def\hodge{\star}

\title{ Topological violation of global symmetries in \\ quantum gravity }

\preprint{TU-1113}

\author{Kazuya Yonekura}
\affiliation{Department of Physics, Tohoku University, Sendai 980-8578, Japan}

\abstract{
We discuss a topological reason why global symmetries are not conserved in quantum gravity, at least when
the symmetry comes from compactification of a higher form symmetry.
The mechanism is purely topological and does not require any explicit breaking term in the UV Lagrangian.
Local current conservation does not imply global charge conservation in a sum over geometries in the path integral. 
We explicitly consider the shift symmetry of an axion-like field which originates from the compactification of a $p$-form gauge field.
Our topological construction is motivated by the brane/black-brane correspondence, brane instantons, 
and an idea that virtual black branes of a simple kind may be realized by surgery on spacetime manifolds.} 

\begin{document}

\maketitle
\section{Introduction}
There is strong evidence that exact global symmetries do not exist in quantum 
gravity (e.g.~\cite{Hawking:1974sw,Banks:1988yz,Coleman:1988cy,Giddings:1988cx,Kallosh:1995hi,ArkaniHamed:2006dz,Banks:2010zn,
Harlow:2018jwu,Harlow:2018tng}). 
One of the arguments is based on Euclidean wormholes~\cite{Coleman:1988cy,Giddings:1988cx}.
Such wormholes may induce terms in the low energy effective theory which violate global symmetries.

There are some conceptual difficulties in the interpretation of the results based on Euclidean wormholes, 
such as the fact that those wormholes
may require baby universes and ensemble of theories (or in other words random couplings) rather than a single theory.
It was not so clear whether we should include them in the gravitational path integral. 
Recently there are several developments such as exact solutions of the JT gravity~\cite{Saad:2019lba}
and replica wormholes~\cite{Penington:2019kki,Almheiri:2019qdq}
which all indicate that Euclidean wormholes are important ingredients
of quantum gravity. These developments, as well as others, 
lead to new insights into violation of global symmetries in quantum gravity, see e.g.~\cite{Harlow:2020bee,Chen:2020ojn,Hsin:2020mfa}.
(See also \cite{Liu:2020sqb,Daus:2020vtf}.) 

However, there are still some mysteries, such as how to interpret the appearance of ensemble of theories~\cite{Saad:2019lba}
(or Coleman's $\alpha$-parameters~\cite{Coleman:1988cy,Giddings:1988cx}).
For example, the problem can be sharpened by considering Euclidean wormholes in AdS spaces
which are known to have a definite CFT dual~\cite{Maldacena:2004rf,ArkaniHamed:2007js}.

In this paper, we study the problem in a broader context of $p$-form symmetries~\cite{Gaiotto:2014kfa}.
(See also \cite{Montero:2017yja,Harlow:2018tng,Rudelius:2020orz} for other discussions.) We argue that 
the relevant geometries for the violation of $p$-form symmetries are obtained by surgery on Euclidean spacetime (and its generalization
which we leave to future work). We will explain surgery and its motivation in quantum gravity in later sections.
Euclidean wormholes are special cases of more general surgery. 
However, the case of $p=0$ (i.e. ordinary symmetry) and $p>0$ are qualitatively different, and
we will see that the situation seems to be much more transparent for $p$-form symmetries with $p>0$
without difficult conceptual issues such as baby universes and random couplings.
Surprisingly, global symmetries are violated even if we have conserved currents, and the violation is caused
just by summing over certain topologically nontrivial configurations.\footnote{While this paper was being completed,
the paper \cite{Hsin:2020mfa} appeared which also discusses global symmetry violation by nontrivial topology
without explicit breaking terms in the Lagrangian.
The difference is that \cite{Hsin:2020mfa} considers 0-form symmetries mainly in the case of ensemble of theories,
while in the present paper we consider $p$-form symmetries for $p>0$ without random couplings.
Then, in our case, global symmetries are violated at the level of amplitudes.}
An important assumption is that those topologically nontrivial configurations contribute in the gravitational path integral.
We will make a little more comments on this point later in the paper. 

In Sec.~\ref{sec:general}, we explain why local current conservation does not imply global charge conservation.
In Sec.~\ref{sec:surgery}, we discuss evidence that virtual black-branes are realized by surgery on spacetime.
Then in Sec.~\ref{sec:shift}, we demonstrate that shift symmetries of axion-like fields which come from the compactification
of $p$-form symmetries are indeed broken.

\section{The general mechanism}\label{sec:general}
Here we would like to discuss why local current conservation does not imply global charge conservation in quantum gravity.
Let us first review how charge conservation is proved in non-gravitational theories
on a fixed background $X = \bR \times Y$, where $\bR$ is the (Euclidean) time direction and $Y$ is space. 
For definiteness, we consider $p$-form $\U(1)$ global symmetries,
although the same conclusion holds for discrete $p$-form symmetries by
more general characterization of symmetry operators as topological operators~\cite{Gaiotto:2014kfa}.\footnote{
The point of view of topological operators make further generalizations possible. 
See e.g.~\cite{Bhardwaj:2017xup,Chang:2018iay,Komargodski:2020mxz}.
There may be more general ``symmetries'' associated to nontrivial bordism classes \cite{McNamara:2019rup,Montero:2020icj},
but we will not discuss them in this paper.}

For a $p$-form $\U(1)$ symmetry, there is a conserved current $J$ which is a $(p+1)$-form operator,
and satisfies the conservation equation
\beq
\d(\star J) =0,
\eeq
where $\star$ is the Hodge star.

On a $D$-dimensional manifold $X = \bR \times Y$ with fixed topology,
we can define a charge as follows. We take a $(D-p-1)$-dimensional submanifold $\Sigma \subset Y$,
and define the charge
$Q(\Sigma_t)$ as
\beq
Q(\Sigma_t) = \int_{\{t\} \times \Sigma} \hodge J, \label{eq:Q}
\eeq 
where $t \in \bR$ is the (Euclidean) time and $\Sigma_t = \{t\} \times \Sigma \subset \bR \times Y$.

Now the conservation of this charge between the initial state at $t=-\infty$
and the final state at $t=+\infty$ can be shown as follows.
By taking $\Gamma= [-\infty, +\infty] \times \Sigma$, we have
\beq
\Sigma_{t=+\infty} - \Sigma_{t=-\infty} = \partial \Gamma. \label{eq:crucial}
\eeq
Therefore, by Stokes theorem we get 
\beq
Q(\Sigma_{t=+\infty})  - Q(\Sigma_{t=-\infty})  = \int_{\Gamma} \d \hodge J=0.
\eeq
In this way, the charge conservation $Q(\Sigma_{t=+\infty})  = Q(\Sigma_{t=-\infty}) $ follows from the current
conservation $\d(\star J)=0$.

In quantum gravity, the above reasoning fails.  
We may take a sum over different topologies, and the partition function or transition amplitude may be given as
\beq
\cZ = \cZ(X) + \cZ(X') + \cZ(X'') + \cdots, \label{eq:QGsum0}
\eeq
where $X= \bR \times Y$, but $X', X''$ and so on, can have complicated topologies.
The only requirement on them is that they have the same asymptotic regions $\{\pm \infty \} \times Y$
(and also the same spatial infinity if $Y$ is not compact) as $X$.\footnote{This requirement is just imposed for the purpose of a clear demonstration
of global symmetry violation in this paper. We can consider more general processes in which the topologies of the initial time
and the final time are different. See \cite{Jafferis:2017tiu} for an interpretation of such cases.}
In other words, we require that 
\beq
\partial X = \partial X' = \partial X'' =\cdots.
\eeq
In the intermediate region, they need not be a product of time $\bR$ and space $Y$.

One might try to show that $Q(\Sigma_{t=+\infty})  = Q(\Sigma_{t=-\infty}) $ by the same reasoning as above.  
For this purpose, the crucial point is to find a submanifold $ \Gamma$ on each of $X, X', X'',\cdots$ whose
boundary is given by $\Sigma_{t=+\infty} - \Sigma_{t=-\infty} $ as in \eqref{eq:crucial}.
If one could find such a $\Gamma$, one would use Stokes theorem to show the charge conservation.
However, as we will show explicitly in Sec.~\ref{sec:shift},
there exists a manifold $X'$ in which we cannot find
any such $\Gamma$. In other words, $\Sigma_{t=-\infty}$ and $\Sigma_{t=+\infty} $
are topologically distinct on that manifold $X'$. In such a configuration,
there is no reason at all that $Q(\Sigma_{t=-\infty}) $ and $Q(\Sigma_{t=+\infty}) $ should be the same.
This is the general reason that a charge is not expected to be conserved in quantum gravity.
We demonstrate it more explicitly in Sec.~\ref{sec:shift}.

The above argument assumes that the relevant manifold $X'$ contributes to the path integral.
In quantum gravity, it is a nontrivial question whether a certain gravitational configuration contributes to
the path integral or not. In particular, we will not show (except for a simple example) that $X'$ satisfies the classical equations of motion.
We have two comments on this point.
First, the results on JT gravity in \cite{Saad:2019lba} encourages us to consider 
gravitational contributions even if they do not satisfy equations of motion, that is, they are not saddle points of the path integral.
Next, our configuration is motivated by some brane considerations in string theory.
We believe that we should sum over brane instantons in the Euclidean path integral.
Our gravitational configuration is motivated by branes under brane/black-brane
correspondence which have led to the discovery of AdS/CFT correspondence; see \cite{Aharony:1999ti} for a review.
Thus, it is reasonable to think that our configuration is relevant for the path integral.

\section{Virtual black branes as surgery}\label{sec:surgery}
Virtual effects of black branes may be incorporated in the Euclidean path integral as 
surgery on spacetime manifolds. Let us first discuss a simple case before explaining surgery in general.

Let us recall Hawking-Page transition~\cite{Hawking:1982dh}. (See also \cite{Witten:1998zw}.)
If we only care about the topology, Hawking-Page transition may be described as follows.
We consider a spacetime with topology $S^1 \times D^{D-1}$, where $D^k$ is a $k$-dimensional disk.
The boundary of this manifold is $S^1 \times S^{D-2}$. Now there is a manifold
$D^2 \times S^{D-2}$ whose boundary is the same as $S^1 \times D^{D-1}$.
The interpretation of it in gravitational theories is as follows. In the geometry $S^1 \times D^{D-1}$,
suppose that we put a black hole at the center $0 \in D^{D-1}$. The world line of the black hole, regarded as a point particle,
wraps $S^1 \times \{0\}$. 
The black hole wrapping $S^1 \times \{0\} \subset S^1 \times D^{D-1}$ is equivalent to the geometry $D^2 \times S^{D-2}$.
This is the topological interpretation of Hawking-Page transition. 
In AdS/CFT, it is important to include both contributions to the path integral
to reproduce confinement/deconfinement phases of CFT~\cite{Witten:1998zw}.

We expect that a natural generalization of the above construction
to a black brane with worldvolume dimension $p$ is as follows. 
Let us consider a spacetime $S^{p} \times D^{D-p}$, and put a brane on $S^p \times \{0\} \subset S^p \times D^{D-p}$.
Then it is equivalent to a smooth geometry $D^{p+1} \times S^{D-p-1}$.
Both manifolds have the same boundary $S^{p} \times S^{D-p-1}$.
In fact, this kind of transition is really realized in string theory in the context of 
deformed/resolved conifolds~\cite{Gopakumar:1998ki,Ooguri:2002gx,Klebanov:2000hb,Maldacena:2000yy,Vafa:2000wi}. 
There, we put D-branes or NS5-branes on $S^p \times \{0\} \subset S^p \times D^{6-p}$ (times $\bR^{4}$),
and by AdS/CFT duality we get a smooth geometry on $D^{p+1} \times S^{5-p}$ (times $\bR^{4}$),
where $p=2\text{~or~}3$.

More generally, when a CFT is put on $S^p$, AdS/CFT itself may be regarded topologically as a duality between
$S^{p} \times D^{D-p}$ with branes wrapped on $S^p \times \{0\}$  and a smooth geometry $D^{p+1} \times S^{D-p-1}$,
where $D^{p+1}$ is a Euclidean AdS (i.e. a hyperbolic space).\footnote{I would like to thank Yuji~Tachikawa
for pointing out this point of view.}
For example, we may consider M5-branes wrapping $S^6$ and then its gravity dual ${\rm AdS}_7 \times S^4$ is topologically 
$D^{p+1} \times S^{D-p-1}$ for $p=6$ and $D=11$.

On a manifold $X$ of general topology, the above transition is described by surgery as follows.
In this paper we assume for simplicity that we have a submanifold $S^p \subset X$ whose normal bundle
is topologically trivial and hence we can take a tubular neighborhood of $S^p$ as $S^p \times D^{D-p} \subset X$.
We wrap a black brane on $S^p \times \{0\} \subset S^p \times D^{D-p} \subset X$.
Then it is equivalent to a new geometry $X'$ in which $S^p \times D^{D-p} $ is replaced by $D^{p+1} \times S^{D-p-1}$
(up to orientation which we mention later).
This is possible since both $S^p \times D^{D-p} $ and $D^{p+1} \times S^{D-p-1}$ have the same boundary $S^p \times S^{D-p-1}$. 
Mathematically, this process of obtaining a new manifold $X'$ from the old one $X$ is called surgery. 
This is done by a local operation in a neighborhood of $S^p$ in $X$.

\section{Violation of shift symmetry of axion-like fields}\label{sec:shift}
Now we would like to demonstrate the violation of global symmetry charge in a more explicit setup.\footnote{The discussions
of this section can be generalized to Chern-Weil symmetries discussed in \cite{Heidenreich:2020pkc}. 
It would be interesting to study the details.}

\subsection{The theory and $p$-form global symmetry} 
We consider the following theory of gravity coupled to a $p$-form gauge field $A$ and possibly other fields.
The Euclidean action $I$ (which appears in the path integral as $\exp(I)$) is
\beq
I =  \frac{1}{16\pi G} \int \d^D x \sqrt{g} R - \frac{2\pi}{2e^2 } \int F \wedge \hodge F + \cdots, \label{eq:Lagr}
\eeq
where $\hodge$ is the Hodge star, $F = \d A$ is the field strength, $D$ is the spacetime dimension, and $G$, $e$ are parameters.
The ellipsis denote possible other matter fields and interactions. 

The gauge field $A$ is normalized as follows. A $p$-dimensional object which has charge $\sq \in \bZ$
under this gauge field has a term in its action given by
\beq
2\pi \i \sq \int A.
\eeq
Then the standard Dirac quantization condition is that $\int F \in \bZ$ on $(p+1)$-cycles.

We call an object which is charged under $A$ as an electric brane. We will consider the case that it is a black brane,
but for the present discussion we need not assume it. 
Suppose that an electric brane is placed at $\{0\} \times \bR^{p} \subset \bR^{D}$.
The Euclidean action contains terms
\beq
I \supset -\frac{2\pi}{2e^2 } \int F \wedge \hodge F +  2\pi \i \sq \int \delta({z}) \wedge A.
\eeq
where ${z} \in \bR^{D-p}$ is the coordinates of $\bR^{D-p}$ which is transverse to the brane, 
and $\delta({z})$ is the delta function $(D-p)$-form localized on ${z}=0$.
The equation of motion is 
\beq
(-1)^p \d \left( e^{-2} \hodge F \right) +  \i \sq(-1)^{p(D-p)}\delta({z})=0
\eeq
We define the electromagnetic dual of $F$ as 
\beq
\widetilde F = \i (-1)^{Dp}e^{-2} \hodge F. \label{eq:FFdual}
\eeq
The factor of the imaginary unit $\i$ is an artifact due to the fact that we are working in Euclidean signature metric
rather than Lorentzian.  
Then, from the above equation we get
\beq
\d \widetilde{F} = \sq \delta({z}) \quad \Longrightarrow \quad \int_{S^{D-p-1}}  \widetilde F = \sq \in \bZ, \label{eq:Eflux}
\eeq
where $S^{D-p-1}$ is a sphere surrounding the electric brane.
This means that the electric brane is a magnetic source of the dual field $\widetilde F$.

The duality between $F$ and $\widetilde F$ can be achieved by the following action 
\beq
I(A, \widetilde F) = 2\pi \int \left( \i (-1)^{D-p-1} \widetilde F \wedge \d A - \frac{e^2}{2} \widetilde F \wedge \hodge \widetilde F   \right). \label{eq:dualityaction}
\eeq
Here we have taken $A$ and $\widetilde F$ as independent variables. If we integrate over $\widetilde F$,
we get back to the original action of $A$ given above with $e^2 \hodge \widetilde F = \i (-1)^{D-p-1} F$ which is equivalent to \eqref{eq:FFdual}.
On the other hand, if we integrate over $A$, we get a dual action for $\widetilde F = \d \widetilde A$
by the following argument. First, the integration over topologically trivial $A$ gives $\d \w F =0$, which is the Bianchi identity.
Next, the sum over nontrivial fluxes of $F = \d A$ sets the fluxes of $\widetilde F$ to be integer-valued.
It can be shown by the formula $\sum_{n \in \bZ} e^{2\pi \i n x} = \sum_{m \in \bZ} \delta(x-m)$ and Poincare duality theorem.
Therefore, after integration over $A$, $\widetilde F$ is described by a dual $(D-p-2)$-form gauge field $\widetilde A$ as $\d \widetilde A = \widetilde F$. 
See \cite{Witten:1995gf} for more detailed discussions
(which was done in the case of $D=4, p=1$ but can be generalized).

The theory has two higher form symmetries~\cite{Gaiotto:2014kfa}.
They are related by duality $F \leftrightarrow \widetilde F$ so we only discuss one of them.
The action of $A$ is invariant under a shift 
\beq
A \to A + \eta \label{eq:Ep-form}
\eeq
where $\eta$ is a closed $p$-form, $\d \eta=0$. 
This is the $p$-form symmetry which is possessed by $A$.
A gauge transformation (rather than a global symmetry transformation) is given by $A \to A + \eta_{\bZ}$,
where $\eta_{\bZ}$ is a closed $p$-form whose integrals on $p$-cycles are integers. 
Thus the actual global symmetry transformations are characterized by elements of $H^p(X; \bR/\bZ)$.
The purpose of this section is to see that this symmetry is broken by the topological reason discussed in Sec.~\ref{sec:general}
even if the Lagrangian does not contain any explicit breaking term.

\subsection{Brane instanton and spacetime surgery}
We make the following set of assumptions on the spacetime geometry $X$.
First, we assume that the theory is compactifed on a closed manifold $M$ of dimension $D-d$ so that the
topology of spacetime is 
\beq
X = \bR^d \times M.
\eeq
We assume that $M$ contains a nontrivial $p$-dimensional cycle $C$ which is topologically $S^p$,
$C \cong S^p$. The cycle $C$ is assumed to generate a free $\bZ$ factor in the homology group $H_p(M;\bZ)$.  

Let $NC$ be the normal bundle to $C$ in $M$, and let $\underline{\bR}^d$ be the trivial rank $d$ bundle
which is the tangent bundle of $\bR^d$ above.
We assume that $NC \oplus \underline{\bR}^d$ is topologically trivial on $C$. 
We remark that this triviality is required only as a real vector bundle as opposed to a complex vector bundle.
For example, let us consider a Calabi-Yau manifold $M$ which contains a complex submanifold $C \cong \mathbb{CP}^1$.
The Calabi-Yau condition requires that the normal bundle $NC$ as a complex bundle has the first Chern class $\int_C c_1(NC) = -2$,
and hence it is nontrivial as a complex vector bundle. However, if we view $NC$ as a real vector bundle by forgetting its complex structure,
its second Stiefel-Whitney class is zero, $\int_C w_2(NC) = -2 = 0 \mod 2$. The second Stiefel-Whitney class is the only obstruction
for a real vector bundle of rank $r >2$ to be trivialized on $\mathbb{CP}^1 \cong S^2$, and hence the normal bundle is trivial
if $r>2$. This is because of the fact that the transition function between the northern hemisphere and the southern hemisphere
is classified by $\pi_1(\SO(r)) \cong \bZ_2$. 
In more general case with $C \cong S^p$, the classification is $\pi_{p-1}(\SO(r))$ (in orientable case).
The relevant rank $r$ in this discussion is that of $NC \oplus \underline{\bR}^d$,
which is $r = D-p$.

Of course, on generic manifolds we can easily find cycles $C$ which do not satisfy the above assumptions.
Sometimes a nontrivial element of $H_p(M;\bZ)$ cannot even be represented by a submanifold.
Normal bundles can also have nontrivial topology. 
We leave it as a future work to understand those more general cases.

By de~Rham theorem, there is a closed $p$-form $\omega$ which we take as a harmonic form such that $\int_C \omega=1$. 
More precisely, we may fix a basis of $H_p(M;\bZ)$ and require $\int \omega =0$ for the basis elements other than $C$.
The low energy theory on $\bR^d$ contains a scalar field $\phi$, an ``axion'', as 
\beq
A = \phi \omega ~~\Longrightarrow~~ \phi = \int_C A.
\eeq 
By gauge transformations, it has periodicity $\phi \sim \phi+1$.
The axion field also has the shift symmetry $\phi \to \phi + \alpha$ for any $\alpha \in \bR$
which comes from the $p$-form symmetry \eqref{eq:Ep-form} by taking $\eta = \alpha \omega$.

Suppose for a moment that there exist ``elementary'' (i.e. not black) branes in the theory.
On the Euclidean spacetime $X = \bR^d \times M$, we can consider a brane instanton of charge $\sq$ by wrapping an electric brane on 
$\{x\} \times C \subset \bR^d \times M =X$.
Its exponentiated action is proportional to $\exp( 2\pi \i \sq \int_{C} A) = \exp(2\pi \i \sq\phi)$. Therefore, the low energy
effective field theory on $\bR^d$ contains a term in the effective action which is proportional to 
\beq
I_{\rm eff} \propto   \int_{\bR^d} \d^d x ( e^{2\pi \i  \sq \phi} + \text{c.c.}). 
\eeq
In particular, it explicitly breaks the shift symmetry $\phi \to \phi + \alpha~( \alpha \in \bR)$ which was present at the classical level.
The only remaining symmetry is $\phi \to \phi+1$, which is a gauge transformation and should not be broken. 

Our purpose here is to see that we can produce the effective action $I_{\rm eff}$ by purely geometric effects 
without considering the electric brane as an ``elementary object''. 
This is expected to be possible by the brane/black-brane correspondence and the idea that a black brane in Euclidean signature space
is realized by surgery, as discussed in Sec.~\ref{sec:surgery}.
At least at the topological level, we argue that it is indeed possible under the assumptions on $C$ made above.

Let us define
\beq
C_{x}=\{ x \} \times C  \subset \bR^d \times M = X .
\eeq
where $x$ is an arbitrary point on $\bR^d$. 
These are the cycles on which the brane instanton would wrap in the above discussion (which is then integrated over $x$).
The normal bundle to $C_x$ in $X$ is $  \underline{\bR}^d \oplus NC$ which
we have assumed to be topologically trivial. We fix one trivialization.
Then we can take a tubular neighborhood $T_x$ of $C_x$ which is topologically $T_x \cong D^{D-p} \times S^p$.
The submanifold $C_x$ lies at the origin of $D^{D-p}$.

Now we perform surgery which is sketched in Sec.~\ref{sec:surgery}. 
First we notice that $T_x$ has a boundary which is topologically $\partial T_x \cong S^{D-p-1} \times S^p$.
This is the same as the boundary of another manifold $S^{D-p-1} \times D^{p+1}$, 
up to an orientation change by $(-1)^{D-p-1}$ since $\partial(S^{D-p-1} \times D^{p+1}) = (-1)^{\dim S^{D-p-1} } S^{D-p-1} \times \partial D^{p+1}$. 
We take a copy of $S^{D-p-1} \times D^{p+1}$ with the orientation change mentioned above,
and denote it as $T'_x$. Notice that $\partial T_x \cong \partial T'_x$.
Surgery is performed by removing $T_x$ from $X$ and then gluing $T'_x$. 
This makes $X$ to another manifold $X'$. 

Notice that the original cycle $C \cong S^p$ is now trivialized near the point $x \in \bR^d$ by the surgery
because we ``filled the inside of $S^p$ by $D^{p+1}$''. 
Also notice that we have created a new nontrivial cycle $S^{D-p-1}$.
On this new cycle, we put an electric flux 
\beq
\int_{S^{D-p-1} } \widetilde F = \sq \in \bZ. \label{eq:fluxsupport}
\eeq
Compare it with \eqref{eq:Eflux}.
We claim that this is the black-brane version of the above electric brane instanton.
The new spacetime $X'$ is completely smooth and it consists of purely the metric $g$ and the gauge field $(A,\w F)$.

Let us study some topological structure of the above flux \eqref{eq:fluxsupport} on the entire manifold $X'$.
We set
\beq
 \widetilde F  =  \sq \eta
\eeq
so that $\int_{S^{D-p-1} }\eta =1$. 
Topologically, the flux given by $\eta $ is equivalent to a delta function $\eta \sim \delta(y-y_0)$ localized on an arbitrary point $y_0 \in S^{D-p-1}$.
In other words, a representative of the Poincare dual of $\eta$ within the local region $T'_x \cong S^{D-p-1} \times D^{p+1}$
is given by $\{y_0 \} \times D^{p+1} \subset T'_x$. The Poincare dual is intuitively regarded as a flux tube of $\widetilde F = \sq \eta$.
(We remark that the discussion here is only at the topological level and the flux of $\w F$ need not be a tube at the dynamical level.)
Its boundary in $T'_x$ is $\{y_0 \} \times S^{p} $, and we can glue 
it to $\bR_+ \times S^{p}$ where $\bR_+ = \{ s \geq 0 \}$ is a copy of the half real line embedded in 
the manifold $X' \setminus T'_x = X \setminus T_x$
in such a way that $s \to \infty $ corresponds to going to infinity on $\bR^{d}$. Topologically, it is irrelevant which direction $\bR_+$ goes to. 
The only point here is that the flux can escape to infinity.

We conclude that topologically $ \eta $ is Poincare-dual to a submanifold $L \subset X'$ with the following properties. 
It is topologically a disk $D^{p+1}$
and the boundary is 
\beq
\partial L = \{\infty\} \times C \cong  S^p, 
\eeq
where $\infty \in S^{d-1} \subset \bR^{d}$ is an arbitrary point at infinity on $\bR^d$. Here $S^{d-1}$ is the boundary at infinity.
This $L$ represents an element of the relative homology group $H_{p+1}(X', \partial X'; \bZ)$
where $\partial X' =\partial X = S^{d-1} \times M$ is the boundary at infinity.
Poincare duality is the isomorphism $H_{p+1}(X', \partial X'; \bZ) \cong H^{D-p-1}(X';\bZ)$ under which
$L$ and $\eta$ correspond to each other.

Now we can see how the geometry generates the term proportional to $e^{2\pi \i \sq \phi}$ in the low energy effective action on $\bR^d$.
First we notice that the original cycle $C$ is now trivialized near $x$ as remarked above by ``filling $S^p$ by $D^{p+1}$'' at the point $x \in \bR^d$.
Then the field $A $ cannot be $\phi \omega$ everywhere on $X'$. We replace it by a configuration $A_\phi$
which goes to $\phi \omega$ at infinity $S^{d-1} \subset \bR^d$, and smoothly goes to zero in the region near $x \in \bR^d$ where we have performed the surgery. We assume that $\phi$ is constant in the neighborhood of the boundary at infinity $S^{d-1} \subset \bR^d$.
Then the field strength $F_\phi = \d A_\phi$ is zero near $S^{d-1}$, and 
it gives an element of cohomology with compact support, $H^{p+1}(X', \partial X'; \bR)$.

We use the action \eqref{eq:dualityaction} in which $A$ and $\widetilde F$ are regarded as independent variables,
and consider the contribution to the path integral coming from the configuration
\beq
A = A_\phi, \qquad \widetilde F = \sq \eta.
\eeq
The relevant part of the action is
\beq
I \supset 2\pi \int \left( \i (-1)^{D-p-1} \widetilde F \wedge \d A - \frac{e^2}{2} \widetilde F \wedge \hodge \widetilde F   \right). 
\eeq
The second term produces a real value for the action.
The first term is the crucial part for our purposes and it is computed as follows. We have seen that $\eta$ is Poincare-dual to the cycle $L$.
After carefully examining the orientation, one gets
\footnote{To check the sign, let us compute it on $T'_x$. We remarked earlier that it is $S^{D-p-1} \times D^{p+1}$ up to an orientation change by $(-1)^{D-p-1}$.
We also have $ \eta = \delta (y-y_0)$ on $S^{D-p-1}$. Thus 
\beq
2\pi \i (-1)^{(D-p-1)} \int_{T'_x}  \sq \eta \wedge F_\phi  = 2\pi \i \sq \int_{S^{D-p-1} \times D^{p+1}} \delta (y-y_0) \wedge F_\phi =2\pi \i\sq \int_{D^{p+1}} F_\phi. 
\eeq
This computation makes sure that the sign in \eqref{eq:ev} is correct. 
 }
\beq
2\pi \i (-1)^{(D-p-1)} \int_{X'}  \sq \eta \wedge F_\phi 
= 2\pi \i \sq \int_L F_\phi = 2\pi \i \sq \int_{\partial L} A_\phi  = 2\pi \i \sq \phi, \label{eq:ev}
\eeq
where we have used the fact that $\partial L$ is at infinity and we can use $A_\phi = \phi \omega$ there.
We have also implicitly used the fact that $F_\phi = \d A_\phi$ defines a cohomology element with compact support as mentioned above,
and computed $\int  \eta \wedge F_\phi$ as a cohomology pairing between $H^{D-p-1}(X';\bR)$ and $H^{p+1}(X', \partial X'; \bR)$.

Therefore, we have reproduced the phase factor $2\pi \i \sq \phi$ which was expected from the brane considerations.
It is a nontrivial question whether the configuration discussed above really contributes to the gravitational path integral.
We regard the brane/black-brane correspondence
and the successful reproduction of the phase $2\pi \i  \sq \phi$ as strong evidence that it really contributes to the path integral.

\subsection{Topological obstruction to charge conservation} \label{sec:obs}

On the manifold $X$ with topology $X = \bR^d \times M$,
the charge $Q(\Sigma_t) = \int_{\{t\} \times \Sigma} \hodge J$ of the $p$-form symmetry current $J$ is conserved
as discussed in Sec.~\ref{sec:general}. 
In the case discussed in this section, 
the $p$-form symmetry current is $\star J = \widetilde F $. 
For $\Sigma$, we may take 
\beq
\Sigma = \bR^{d-1} \times \widetilde C,
\eeq
where $\widetilde C$ is a submanifold on $M$ which is Poincare-dual to $\omega$ on $M$.
In particular the intersection number between $C$ and $\widetilde C$ is 1 up to sign,
since Poincare-duality implies $\int_{C \cap \w C} 1= \int_C \omega =1$.
$\bR^{d-1}$ is a ``time slice'' on $\bR^d$ though we are working with Euclidean signature metric. 

In quantum gravity we may take a sum over geometries with different topologies, 
and the partition function or transition amplitude may be given as
\beq
\cZ = \cZ(X) + \cZ(X') + \cdots. \label{eq:QGsum}
\eeq
We claimed in Sec.~\ref{sec:general} that for some $X'$, there is no $\Gamma$
such that $\partial  \Gamma = \Sigma_{t=+\infty} - \Sigma_{t=-\infty}$.
Here we prove the non-existence of $\Gamma$
for the $X'$ which is obtained from $X = \bR^d \times M$ by the surgery
as constructed above.

Before proving it, let us make one remark.
In the conservation argument given in Sec.~\ref{sec:general}, we have implicitly neglected spatial infinity on $\bR^{d-1}$.
More precisely, including the spatial infinity, we replace $ \Sigma_{t=+\infty} - \Sigma_{t=-\infty}$ by
$ S^{d-1} \times \widetilde C$, where $S^{d-1}$ is the sphere at infinity on $\bR^d$.
Then the question is whether we can find a submanifold $\Gamma$ such that $\partial \Gamma = S^{d-1} \times \w C$.

Recall that $\w C$ is defined to be the Poincare dual of $\omega$.
Let $ A_{\phi=1}$ be a $p$-form, defined in a similar way as $A_\phi$, which is $\omega$ at the spatial infinity
and goes to zero in the region near $T'_x$.
Then we see that 
\beq
\int_{S^{d-1} \times \widetilde C}  \eta = \int_{S^{d-1} \times M }   \eta \wedge \omega 
= (-1)^{D-p-1}  \int_{X' }   \eta \wedge \d A_{\phi=1} =1  ,\label{eq:obstruction}
\eeq
where we have used \eqref{eq:ev} with $\phi$ set to 1. 
If there were $\Gamma$ such that $\partial \Gamma = S^{d-1} \times \widetilde C $, we would get
$\int_{S^{d-1} \times \widetilde C} \eta = \int_{\Gamma} \d \eta=0$,
a contradiction. Therefore, there is no such $\Gamma$. 

The above argument shows that the flux $\widetilde F = \sq \eta$ is precisely the obstruction to the existence of $\Gamma$ on $X'$ such that 
$\partial \Gamma = S^{d-1} \times \widetilde C$.
The flux $\widetilde F$ have played an important role in the generation of the phase $2\pi \i \sq \phi$ in the above discussions,
and it also gives an obstruction to the existence of $\Gamma$.
We conclude that $Q(\Sigma_{t=-\infty})$ and $Q(\Sigma_{t=+\infty})$ need not be the same on $X'$,
and the charge conservation is violated in the sum \eqref{eq:QGsum}.

\subsection{Example}\label{sec:example}
There is a simple example which is actually a saddle point of the path integral at the classical level.\footnote{
This example is subtle. 
We neglect possible quantum mechanical instabilities, such as the one discussed in \cite{Witten:1981gj}. 
In fact, our configuration is similar to that of \cite{Witten:1981gj},  
but we want to interpret it as a loop of a ``virtual charged particle'' as discussed later. Also, 
it is not so certain how to think about it since it can also happen in string theory~\cite{GarciaEtxebarria:2020xsr}. 
We leave it as a future work to better understand it.
} Let us discuss this example. 
Similar discussions on global symmetry violation have also been given in Sec.~4.3 of \cite{Montero:2017yja}. 

We consider $D=4$, $p=1$, $d=3$ and $M=C=S^1$. Thus the manifold $X$ is $X = \bR^3 \times S^1$ where $S^1$ is the internal manifold in the compactification. 
In this situation, we can consider the Euclidean version of the usual Reissner-Nordstr\"om black hole
going around $S^1$. Only the interpretation is different. Usually, $S^1$ is considered as 
a Wick-rotated time direction and the black hole solution is considered to represent the thermodynamic state
of the black hole. Instead of that interpretation, in our context, we interpret $S^1$ as one of the space directions,
and the time direction is included in $\bR^3$. We sum over geometries $X, X', X''.\cdots,$
with no virtual black hole, one virtual black hole, two virtual black holes etc. in the path integral.
The shift symmetry of the axion field $\phi = \int_{S^1} A$ is violated.

More explicitly, the situation is as follows. First let us describe the topology. 
By considering a ``sphere at infinity'' $S^2 \subset \bR^3$ as the boundary of $\bR^3$, we may regard $\bR^3$ topologically  as a three-dimensional disk $D^3$.
Thus we regard $X = \bR^3 \times S^1$ topologically as $X \cong D^3 \times S^1$. By performing surgery near $\{0\} \times S^1 \subset D^3 \times S^1$,
we get $X' \cong S^2 \times D^2$. Both $X$ and $X'$ have the same asymptotic boundaries $\partial (D^3 \times S^1) \cong \partial (S^2 \times D^2) \cong S^2 \times S^1$.
We also put a flux $\int_{S^2} \w F = \sq$ on the $S^2$ of $X' \cong S^2 \times D^2$. This construction of $X'$ is an explicit example of the general construction described in the previous subsections.

The geometry of $X'$ is that of the Reissner-Nordstr\"om black hole and is given by
\beq
\d s^2 &= \frac{(r-r_+)(r-r_-)}{r^2}  \d \tau^2 +  \frac{r^2}  {(r-r_+)(r-r_-)} \d r^2 + r^2(\d \theta^2 + \sin^2\theta \d \phi^2), \nonumber \\
\w F &= \frac{\sq}{4\pi} \sin \theta \d \theta \wedge \d \phi.
\eeq
The parameters $r_+$ and $r_-$ are given in terms of the radius $R$ of $S^1 = \{ \tau; \, \tau \sim \tau+2\pi R \}$ as
\beq
R = \frac{2 r_+^2}{r_+ - r_-}, \qquad r_+ r_- = \frac{1}{2} G e^2 \sq^2.\label{eq:rpm}
\eeq
The first equation is required by the condition that $(r,\tau)$ form a smooth disk $D^2$ in the region $r \geq r_+$,
where $r=r_+$ is the center $\{0\}$ of $D^2$.
The second one comes from Einstein equations.

The coordinates $(r, \tau)$ give the disk $D^2$ of $X' \cong S^2 \times D^2$. More precisely, $(\sqrt{r-r_+},  \tau/R)$ are the polar coordinates of it. 
 The coordinates $(\theta, \phi)$ give the sphere $S^2$ of $X' \cong S^2 \times D^2$.
In the region $r \to \infty$, both $X$ and $X'$ have the same asymptotic regions described by the metric $\d s^2 \to   \d \tau^2 +   \d r^2 + r^2(\d \theta^2 + \sin^2\theta \d \phi^2)$.
In this asymptotic region, we may take Euclidean coordinates
\beq
(t, x, y,\tau) =  (r\cos\theta, r\sin\theta \cos\phi, r\sin\theta \sin\phi, \tau)
\eeq
and we may regard $t$ as the (Euclidean) time coordinate. The cycles $\Sigma_{t=\pm \infty}$ on which we define initial and final charges are taken to be
$\Sigma_{t=\pm \infty} = \{\pm \infty \} \times \bR^2 \times \{\mathrm{pt} \}$, where $\mathrm{pt} \in S^1$ is a point on $S^1$. One can see that there
is no $\Gamma$ on $X'$ such that $\partial \Gamma = \Sigma_{t=+ \infty} - \Sigma_{t=\pm \infty}$ 
(or more precisely $\partial \Gamma = S^2 \times \{\mathrm{pt} \}$ where $S^2$ is the sphere at infinity)
as was proved more generally in Sec.~\ref{sec:obs}.

To compute the gravitational action $I$ of the above geometry, we borrow a formula from the thermodynamic interpretation
of the above configuration~\cite{Gibbons:1976ue}. If we interpret $M = S^1$ as the Euclidean time direction, the action $I$
is proportional to the free energy of the black hole,
\beq
\frac{I}{2\pi R} = - E + TS + \sq \mu,
\eeq
where the energy $E$, the temperature $T$, the entropy $S$ and the chemical potential $\mu$ are given by
\beq
E= \frac{(r_+ + r_-)}{2G}, \qquad T = \frac{1}{2\pi R}, \qquad S = \frac{4\pi r_+^2}{4G}, \qquad \mu = \frac{\i}{ R} \phi
\eeq
Therefore, the gravitational path integral is proportional to
\beq
\exp(I) = e^{S} \exp( - 2\pi R E+ 2\pi i \sq \phi ).\label{eq:BHaction}
\eeq
The amount of the shift symmetry breaking of $\phi = \int_{S^1} A$ is proportional to this quantity. 

In our case, $M=S^1$ is not a Euclidean time direction, but one of the spatial directions with a fixed radius $R$.
For large enough $R \gg \sqrt{G e^2 \sq^2}$, there are two solutions to \eqref{eq:rpm}, and one of them is given by
$r_+ \simeq r_- \simeq \sqrt{G e^2 \sq^2/2}$ and $r_+ - r_- \simeq G e^2 \sq^2/R$ which is near extremal.
For this solution, the formula \eqref{eq:BHaction} may be interpreted as follows. 
Let us regard the black hole as a particle.
The usual action for a relativistic particle of mass $m$ and charge $\sq$ is 
\beq
I_{\rm particle} = -m \int \d s + 2\pi \i \sq \int A.
\eeq
When a virtual particle goes around the $M=S^1$, the first term gives $-2\pi R m$ with $m=E$, where $\int \d s =2\pi R$,
and the second term gives $2\pi i \sq \phi$.
Moreover, the particle has internal states, and the number of states which contribute to the virtual particle loop around $M=S^1$
is of order $e^{S}$.\footnote{We remark that the total number of particle states can be infinity.
Only the number of states which give significant contributions to the path integral is of order $e^{S}$.
For example, states whose energy $E'$ is significantly larger than the averaged energy $E$
do not contribute if $E'-E \gg R^{-1}$ since their contributions are exponentially suppressed.
}
This virtual particle interpretation gives the action \eqref{eq:BHaction}.

\section{Discussions}
\subsection{Brief summary of the case $p>0$}
In this paper we have discussed that global symmetry charges are not expected to be conserved in quantum gravity.
We sum over topologies
\beq
X+X' + X'' + \cdots.\label{eq:summary_sum}
\eeq
For definiteness, we require that their boundary manifolds are topologically the same,
\beq
\partial X = \partial X' = \partial X'' =\cdots. \label{eq:summary_same}
\eeq
In particular, the initial and final states are assumed to have the common topology in all $X,X',X''$ and so on.\footnote{This assumption is not really a requirement of quantum gravity,
but is made to clarify our charge violation mechanism.}

The standard proof of global charge conservation is done by using the Stokes theorem.
Namely, we take a cycle $\Sigma_{t =\pm \infty}$ at the initial time $t = -\infty$ and the final time $t=+\infty$,
and define the global charges at $t=\pm \infty$ as $Q(\Sigma_{\pm \infty}) = \int_{\{t=\pm \infty \} \times \Sigma} \star J$.
Then we can show that $Q(\Sigma_{+ \infty}) - Q(\Sigma_{- \infty}) = \int_{\Gamma} \d(\star J) =0$ if there is some appropriate $\Gamma$
such that $\partial \Gamma = \Sigma_{+ \infty} - \Sigma_{- \infty}$.

We have pointed out that the above proof fails in quantum gravity in which we take a sum as in \eqref{eq:summary_sum}.
By our simplifying assumption \eqref{eq:summary_same} 
that all $X, X', X''$ and so on have the common initial and final state topology, the cycles $\Sigma_{t =\pm \infty}$ are well-defined.
However, in some $X'$, we fail to find a $\Gamma$ with the above property.
Thus, from the beginning, there was no proof of charge conservation. 

In the case of $0$-form symmetries, the above mechanism involves some conceptual issues such as two disjoint universes as we discuss later.
However, the charge violation is very clear in the case of higher form global symmetries which are compactified to 0-form symmetries. 

We consider a $D$-dimensional theory with a $p$-form global symmetry, and compactify the theory down to $d$-dimensions as
$\bR^d \times M$, where $M$ is an internal manifold. Explicitly, we have studied the case of a $p$-form gauge field which has a $p$-form shift symmetry.
Depending on the topology of $M$, we get a 0-form axion-like field with a shift symmetry in the low energy effective
theory on $\bR^d$. 
In quantum gravity, the shift symmetry is broken and in generic situations (without supersymmetry etc. so that there is no fermion zero modes), 
the axion-like field gets a potential energy.

We emphasize that the generation of the potential energy of the axion-like field
is a clear physical effect, irrespective of conceptual interpretations of our results:
\begin{enumerate}
\item In the case of quantum field theory in which we do not sum over topologies, 
the axion-like field is a Goldstone boson in the low energy theory on $\bR^d$, and 
in particular it is exactly massless. This is due to the shift symmetry which is ensured by the higher dimensional $p$-form symmetry.
\item In the case of quantum gravity in which we sum over some relevant topologies, the axion-like field is no longer massless in generic situations (unless
it is forced to be massless by other reasons such as supersymmetry.) In any case, the shift symmetry is broken in the low energy theory on $\bR^d$.
\end{enumerate}
There is no conceptual issue in these statements, since e.g. the mass of the axion-like field is a physical observable. 
We can even give an order estimate of the size of the shift symmetry breaking as in the example in Sec.~\ref{sec:example}.
We have produced the mass term (or more general shift symmetry breaking terms) without introducing any explicit breaking term in the UV Lagrangian
of the higher dimensional theory. This is a clear difference from the case of quantum field theory.

\subsection{Speculations on the case $p=0$}
The construction described in Sec.~\ref{sec:shift} requires
a cycle $C$ which is topologically $S^p$, and represents a nontrivial 
element of the homology $H_p(M; \bZ)$. 
There is a qualitative difference between the cases $p>0$ and $p=0$.

For $p>0$, a sphere $S^p$ is connected. On the other hand, for $p=0$
a sphere $S^0$ consists of two points. Moreover, the fact that $C \cong S^0$ needs to be a
nontrivial element of $H_0(M;\bZ)$ implies that $M$ needs to have at least two components,
$M = M_1 \sqcup M_2$, where $\sqcup$ means disjoint union. 
One point of $S^0$ is on $M_1$, and the other is on $M_2$.
Thus we can think of $X_1 = \bR^d \times M_1$ as ``one universe'' and $X_2 = \bR^d \times M_2$ as ``another universe'',
and they are disconnected from each other. 
Then our virtual black brane for $p=0$ is exactly the Euclidean axionic wormhole~\cite{Giddings:1987cg},
and in the current case it is topologically the connected sum $X_1 \# X_2$ of the two universes.
Such wormholes are argued to violate usual (i.e. 0-form) global symmetries~\cite{Coleman:1988cy,Giddings:1988cx}.
In our situation, the violation happens by a flow of global symmetry charge from one universe to the other through the wormhole.

In the case $p=0$, the wormhole poses conceptual questions about its interpretation. 
In particular, \cite{Maldacena:2004rf,ArkaniHamed:2007js} (see also \cite{Betzios:2019rds}) have presented a sharp problem in AdS/CFT
by constructing wormholes in AdS where there is a single CFT dual.
The problem becomes even more severe in the light of the discovery that 
off-shell configurations which are not saddle points can contribute to the path integral~\cite{Saad:2019lba}.

From the point of view of the brane/black-brane correspondence, 
the Euclidean wormhole may be considered to be equivalent to
two $(-1)$-branes, or more precisely one $(-1)$-brane and one anti-$(-1)$-brane, since $S^0$ consists of two points. 
Then we conjecture that there must be a $(-1)$-brane in a theory
which is really a single unitary theory in a single universe without random couplings.
A very related proposal has also been given in \cite{Marolf:2020xie,McNamara:2020uza} from a different (but related) viewpoint.

The reason for the necessity of a $(-1)$-brane is as follows.\footnote{
We may also call $(-1)$-branes as ``half-wormholes''. See \cite{Saad:2021rcu} for recent work on half-wormholes.}
Most of the problems about Euclidean wormholes
are due to the fact that they connect two different points in spacetime in a nonlocal way.
However, let us replace a Euclidean wormhole by a pair of a $(-1)$-brane and an anti-$(-1)$-brane.
Then, there is no reason to only sum over configurations in which the numbers of $(-1)$-branes
and anti-$(-1)$-branes are the same. We can consider configurations with independent numbers 
$N_+$ and $N_-$ of $(-1)$-branes and anti-$(-1)$-branes, respectively, and sum over $N_+$ and $N_-$
independently. Then we do not get nonlocal interactions between two points. 
A pair of a $(-1)$-brane and an anti-$(-1)$-brane 
must induce the same symmetry violation as a Euclidean wormhole,
and hence $(-1)$-branes must transform nontrivially under the global symmetry and then lead to its breaking.

In fact, we can consider an analogous problem in non-gravitational theories.
Let us consider a four dimensional gauge theory like QCD on $\bR^4$. 
We assume that the theory is higgsed at a high energy scale so that instanton gas approximation is valid. 
If we only sum over configurations with total instanton number 0, we get a problem of nonlocal interactions.
For example, the leading nontrivial contribution comes from an instanton at a point $x \in \bR^4$
and an anti-instanton at $y \in \bR^4$ so that the total instanton number is zero. 
Then we get an effective nonlocal interaction of the form
\beq
\int \d^4x \d^4 y\, O(x) \overline{O}(y) +\cdots
\eeq
where $O(x)$ is produced by an instanton (such as an 't~Hooft vertex~\cite{tHooft:1976rip}), and $\overline{O}(y)$ is produced by
an anti-instanton. However, if we sum over all configurations of arbitrary instanton and anti-instanton numbers, we get
\beq
\exp\left( \int \d^4x \, O(x) \right) \exp\left( \int \d^4y \, \overline{O}(y) \right) = \exp\left( \int \d^4x \,  [O(x) +\overline{O}(x)] \right).
\eeq
This is the reason that we sum over all configurations of instantons on $\bR^4$.
See \cite{Weinberg:1996kr} for more discussions.\footnote{
On the other hand, on a spacetime of compact topology rather than $\bR^4$,
it is possible to impose some constraints on instanton numbers; see e.g. \cite{Seiberg:2010qd,Tanizaki:2019rbk,Pantev:2005rh,Pantev:2005zs,Pantev:2005wj}.
The difference between noncompact and compact cases is due to some degrees of freedom
which is frozen in noncompact spaces but is summed over in compact spaces. For example,
if a discrete 0-form $\bZ_k$ symmetry is spontaneously broken, we pick a single vacuum on $\bR^4$ to satisfy cluster decomposition,
but we sum over all vacua on noncompact spaces.
} In this context, the analog of Coleman's $\alpha$-parameters may be the $\theta$-angle associated to the instanton numbers.
See also \cite{Ohmori:2021fms} for an interesting field theoretical analog of replica wormholes.

The existence of $(-1)$-branes does not immediately solve all the problems about Euclidean wormholes
due to more detailed problems beyond topology, such as the numerical value of the action of the Euclidean wormhole~\cite{ArkaniHamed:2007js}.
But at least at the topological level, the existence of $(-1)$-branes seems to be a necessary ingredient.

In the case of configurations studied in this paper with $p>0$, we performed surgery only near a point $x \in \bR^d$ in a single universe.
Thus we do not get any nonlocal interactions in the low energy theory on $\bR^d$.
This is the conceptual simplicity of our setup for $p>0$.
In fact, from the point of view of a low energy observer on $\bR^d$, our configurations look like a $(-1)$-brane. 
Thus we can see global symmetry violation in a clean way without difficult conceptual issues. 

\acknowledgments

I am very grateful to Yuji Tachikawa for stimulating discussions which have led to this work.
KY is in part supported by JSPS KAKENHI Grant-in-Aid (Wakate-B), No.17K14265.


\bibliographystyle{JHEP}
\bibliography{ref}

\providecommand{\href}[2]{#2}\begingroup\raggedright\begin{thebibliography}{10}

\bibitem{Hawking:1974sw}
S.~Hawking, {\it {Particle Creation by Black Holes}},  {\em Commun. Math.
  Phys.} {\bf 43} (1975) 199--220. [Erratum: Commun.Math.Phys. 46, 206 (1976)].

\bibitem{Banks:1988yz}
T.~Banks and L.~J. Dixon, {\it {Constraints on String Vacua with Space-Time
  Supersymmetry}},  {\em Nucl. Phys. B} {\bf 307} (1988) 93--108.

\bibitem{Coleman:1988cy}
S.~R. Coleman, {\it {Black Holes as Red Herrings: Topological Fluctuations and
  the Loss of Quantum Coherence}},  {\em Nucl. Phys. B} {\bf 307} (1988)
  867--882.

\bibitem{Giddings:1988cx}
S.~B. Giddings and A.~Strominger, {\it {Loss of Incoherence and Determination
  of Coupling Constants in Quantum Gravity}},  {\em Nucl. Phys. B} {\bf 307}
  (1988) 854--866.

\bibitem{Kallosh:1995hi}
R.~Kallosh, A.~D. Linde, D.~A. Linde, and L.~Susskind, {\it {Gravity and global
  symmetries}},  {\em Phys. Rev. D} {\bf 52} (1995) 912--935,
  [\href{http://arxiv.org/abs/hep-th/9502069}{{\tt hep-th/9502069}}].

\bibitem{ArkaniHamed:2006dz}
N.~Arkani-Hamed, L.~Motl, A.~Nicolis, and C.~Vafa, {\it {The String landscape,
  black holes and gravity as the weakest force}},  {\em JHEP} {\bf 06} (2007)
  060, [\href{http://arxiv.org/abs/hep-th/0601001}{{\tt hep-th/0601001}}].

\bibitem{Banks:2010zn}
T.~Banks and N.~Seiberg, {\it {Symmetries and Strings in Field Theory and
  Gravity}},  {\em Phys. Rev. D} {\bf 83} (2011) 084019,
  [\href{http://arxiv.org/abs/1011.5120}{{\tt arXiv:1011.5120}}].

\bibitem{Harlow:2018jwu}
D.~Harlow and H.~Ooguri, {\it {Constraints on Symmetries from Holography}},
  {\em Phys. Rev. Lett.} {\bf 122} (2019), no.~19 191601,
  [\href{http://arxiv.org/abs/1810.05337}{{\tt arXiv:1810.05337}}].

\bibitem{Harlow:2018tng}
D.~Harlow and H.~Ooguri, {\it {Symmetries in quantum field theory and quantum
  gravity}},  \href{http://arxiv.org/abs/1810.05338}{{\tt arXiv:1810.05338}}.

\bibitem{Saad:2019lba}
P.~Saad, S.~H. Shenker, and D.~Stanford, {\it {JT gravity as a matrix
  integral}},  \href{http://arxiv.org/abs/1903.11115}{{\tt arXiv:1903.11115}}.

\bibitem{Penington:2019kki}
G.~Penington, S.~H. Shenker, D.~Stanford, and Z.~Yang, {\it {Replica wormholes
  and the black hole interior}},  \href{http://arxiv.org/abs/1911.11977}{{\tt
  arXiv:1911.11977}}.

\bibitem{Almheiri:2019qdq}
A.~Almheiri, T.~Hartman, J.~Maldacena, E.~Shaghoulian, and A.~Tajdini, {\it
  {Replica Wormholes and the Entropy of Hawking Radiation}},  {\em JHEP} {\bf
  05} (2020) 013, [\href{http://arxiv.org/abs/1911.12333}{{\tt
  arXiv:1911.12333}}].

\bibitem{Harlow:2020bee}
D.~Harlow and E.~Shaghoulian, {\it {Global symmetry, Euclidean gravity, and the
  black hole information problem}},
  \href{http://arxiv.org/abs/2010.10539}{{\tt arXiv:2010.10539}}.

\bibitem{Chen:2020ojn}
Y.~Chen and H.~W. Lin, {\it {Signatures of global symmetry violation in
  relative entropies and replica wormholes}},
  \href{http://arxiv.org/abs/2011.06005}{{\tt arXiv:2011.06005}}.

\bibitem{Hsin:2020mfa}
P.-S. Hsin, L.~V. Iliesiu, and Z.~Yang, {\it {A violation of global symmetries
  from replica wormholes and the fate of black hole remnants}},
  \href{http://arxiv.org/abs/2011.09444}{{\tt arXiv:2011.09444}}.

\bibitem{Liu:2020sqb}
J.~Liu, {\it {Scrambling and decoding the charged quantum information}},  {\em
  Phys. Rev. Res.} {\bf 2} (2020) 043164,
  [\href{http://arxiv.org/abs/2003.11425}{{\tt arXiv:2003.11425}}].

\bibitem{Daus:2020vtf}
T.~Daus, A.~Hebecker, S.~Leonhardt, and J.~March-Russell, {\it {Towards a
  Swampland Global Symmetry Conjecture using weak gravity}},  {\em Nucl. Phys.
  B} {\bf 960} (2020) 115167, [\href{http://arxiv.org/abs/2002.02456}{{\tt
  arXiv:2002.02456}}].

\bibitem{Maldacena:2004rf}
J.~M. Maldacena and L.~Maoz, {\it {Wormholes in AdS}},  {\em JHEP} {\bf 02}
  (2004) 053, [\href{http://arxiv.org/abs/hep-th/0401024}{{\tt
  hep-th/0401024}}].

\bibitem{ArkaniHamed:2007js}
N.~Arkani-Hamed, J.~Orgera, and J.~Polchinski, {\it {Euclidean wormholes in
  string theory}},  {\em JHEP} {\bf 12} (2007) 018,
  [\href{http://arxiv.org/abs/0705.2768}{{\tt arXiv:0705.2768}}].

\bibitem{Gaiotto:2014kfa}
D.~Gaiotto, A.~Kapustin, N.~Seiberg, and B.~Willett, {\it {Generalized Global
  Symmetries}},  {\em JHEP} {\bf 02} (2015) 172,
  [\href{http://arxiv.org/abs/1412.5148}{{\tt arXiv:1412.5148}}].

\bibitem{Montero:2017yja}
M.~Montero, A.~M. Uranga, and I.~Valenzuela, {\it {A Chern-Simons Pandemic}},
  {\em JHEP} {\bf 07} (2017) 123, [\href{http://arxiv.org/abs/1702.06147}{{\tt
  arXiv:1702.06147}}].

\bibitem{Rudelius:2020orz}
T.~Rudelius and S.-H. Shao, {\it {Topological Operators and Completeness of
  Spectrum in Discrete Gauge Theories}},
  \href{http://arxiv.org/abs/2006.10052}{{\tt arXiv:2006.10052}}.

\bibitem{Bhardwaj:2017xup}
L.~Bhardwaj and Y.~Tachikawa, {\it {On finite symmetries and their gauging in
  two dimensions}},  {\em JHEP} {\bf 03} (2018) 189,
  [\href{http://arxiv.org/abs/1704.02330}{{\tt arXiv:1704.02330}}].

\bibitem{Chang:2018iay}
C.-M. Chang, Y.-H. Lin, S.-H. Shao, Y.~Wang, and X.~Yin, {\it {Topological
  Defect Lines and Renormalization Group Flows in Two Dimensions}},  {\em JHEP}
  {\bf 01} (2019) 026, [\href{http://arxiv.org/abs/1802.04445}{{\tt
  arXiv:1802.04445}}].

\bibitem{Komargodski:2020mxz}
Z.~Komargodski, K.~Ohmori, K.~Roumpedakis, and S.~Seifnashri, {\it {Symmetries
  and Strings of Adjoint QCD${}_2$}},
  \href{http://arxiv.org/abs/2008.07567}{{\tt arXiv:2008.07567}}.

\bibitem{McNamara:2019rup}
J.~McNamara and C.~Vafa, {\it {Cobordism Classes and the Swampland}},
  \href{http://arxiv.org/abs/1909.10355}{{\tt arXiv:1909.10355}}.

\bibitem{Montero:2020icj}
M.~Montero and C.~Vafa, {\it {Cobordism Conjecture, Anomalies, and the String
  Lamppost Principle}},  \href{http://arxiv.org/abs/2008.11729}{{\tt
  arXiv:2008.11729}}.

\bibitem{Jafferis:2017tiu}
D.~L. Jafferis, {\it {Bulk reconstruction and the Hartle-Hawking
  wavefunction}},  \href{http://arxiv.org/abs/1703.01519}{{\tt
  arXiv:1703.01519}}.

\bibitem{Aharony:1999ti}
O.~Aharony, S.~S. Gubser, J.~M. Maldacena, H.~Ooguri, and Y.~Oz, {\it {Large N
  field theories, string theory and gravity}},  {\em Phys. Rept.} {\bf 323}
  (2000) 183--386, [\href{http://arxiv.org/abs/hep-th/9905111}{{\tt
  hep-th/9905111}}].

\bibitem{Hawking:1982dh}
S.~Hawking and D.~N. Page, {\it {Thermodynamics of Black Holes in anti-De
  Sitter Space}},  {\em Commun. Math. Phys.} {\bf 87} (1983) 577.

\bibitem{Witten:1998zw}
E.~Witten, {\it {Anti-de Sitter space, thermal phase transition, and
  confinement in gauge theories}},  {\em Adv. Theor. Math. Phys.} {\bf 2}
  (1998) 505--532, [\href{http://arxiv.org/abs/hep-th/9803131}{{\tt
  hep-th/9803131}}].

\bibitem{Gopakumar:1998ki}
R.~Gopakumar and C.~Vafa, {\it {On the gauge theory / geometry
  correspondence}},  {\em AMS/IP Stud. Adv. Math.} {\bf 23} (2001) 45--63,
  [\href{http://arxiv.org/abs/hep-th/9811131}{{\tt hep-th/9811131}}].

\bibitem{Ooguri:2002gx}
H.~Ooguri and C.~Vafa, {\it {World sheet derivation of a large N duality}},
  {\em Nucl. Phys. B} {\bf 641} (2002) 3--34,
  [\href{http://arxiv.org/abs/hep-th/0205297}{{\tt hep-th/0205297}}].

\bibitem{Klebanov:2000hb}
I.~R. Klebanov and M.~J. Strassler, {\it {Supergravity and a confining gauge
  theory: Duality cascades and chi SB resolution of naked singularities}},
  {\em JHEP} {\bf 08} (2000) 052,
  [\href{http://arxiv.org/abs/hep-th/0007191}{{\tt hep-th/0007191}}].

\bibitem{Maldacena:2000yy}
J.~M. Maldacena and C.~Nunez, {\it {Towards the large N limit of pure N=1
  superYang-Mills}},  {\em Phys. Rev. Lett.} {\bf 86} (2001) 588--591,
  [\href{http://arxiv.org/abs/hep-th/0008001}{{\tt hep-th/0008001}}].

\bibitem{Vafa:2000wi}
C.~Vafa, {\it {Superstrings and topological strings at large N}},  {\em J.
  Math. Phys.} {\bf 42} (2001) 2798--2817,
  [\href{http://arxiv.org/abs/hep-th/0008142}{{\tt hep-th/0008142}}].

\bibitem{Heidenreich:2020pkc}
B.~Heidenreich, J.~McNamara, M.~Montero, M.~Reece, T.~Rudelius, and
  I.~Valenzuela, {\it {Chern-Weil Global Symmetries and How Quantum Gravity
  Avoids Them}},  \href{http://arxiv.org/abs/2012.00009}{{\tt
  arXiv:2012.00009}}.

\bibitem{Witten:1995gf}
E.~Witten, {\it {On S duality in Abelian gauge theory}},  {\em Selecta Math.}
  {\bf 1} (1995) 383, [\href{http://arxiv.org/abs/hep-th/9505186}{{\tt
  hep-th/9505186}}].

\bibitem{Witten:1981gj}
E.~Witten, {\it {Instability of the Kaluza-Klein Vacuum}},  {\em Nucl. Phys. B}
  {\bf 195} (1982) 481--492.

\bibitem{GarciaEtxebarria:2020xsr}
I.~Garcia~Etxebarria, M.~Montero, K.~Sousa, and I.~Valenzuela, {\it {Nothing is
  certain in string compactifications}},
  \href{http://arxiv.org/abs/2005.06494}{{\tt arXiv:2005.06494}}.

\bibitem{Gibbons:1976ue}
G.~Gibbons and S.~Hawking, {\it {Action Integrals and Partition Functions in
  Quantum Gravity}},  {\em Phys. Rev. D} {\bf 15} (1977) 2752--2756.

\bibitem{Giddings:1987cg}
S.~B. Giddings and A.~Strominger, {\it {Axion Induced Topology Change in
  Quantum Gravity and String Theory}},  {\em Nucl. Phys. B} {\bf 306} (1988)
  890--907.

\bibitem{Betzios:2019rds}
P.~Betzios, E.~Kiritsis, and O.~Papadoulaki, {\it {Euclidean Wormholes and
  Holography}},  {\em JHEP} {\bf 06} (2019) 042,
  [\href{http://arxiv.org/abs/1903.05658}{{\tt arXiv:1903.05658}}].

\bibitem{Marolf:2020xie}
D.~Marolf and H.~Maxfield, {\it {Transcending the ensemble: baby universes,
  spacetime wormholes, and the order and disorder of black hole information}},
  {\em JHEP} {\bf 08} (2020) 044, [\href{http://arxiv.org/abs/2002.08950}{{\tt
  arXiv:2002.08950}}].

\bibitem{McNamara:2020uza}
J.~McNamara and C.~Vafa, {\it {Baby Universes, Holography, and the Swampland}},
   \href{http://arxiv.org/abs/2004.06738}{{\tt arXiv:2004.06738}}.

\bibitem{Saad:2021rcu}
P.~Saad, S.~H. Shenker, D.~Stanford, and S.~Yao, {\it {Wormholes without
  averaging}},  \href{http://arxiv.org/abs/2103.16754}{{\tt arXiv:2103.16754}}.

\bibitem{tHooft:1976rip}
G.~'t~Hooft, {\it {Symmetry Breaking Through Bell-Jackiw Anomalies}},  {\em
  Phys. Rev. Lett.} {\bf 37} (1976) 8--11.

\bibitem{Weinberg:1996kr}
S.~Weinberg, {\em {The quantum theory of fields. Vol. 2: Modern applications}}.
\newblock Cambridge University Press, 8, 2013.

\bibitem{Seiberg:2010qd}
N.~Seiberg, {\it {Modifying the Sum Over Topological Sectors and Constraints on
  Supergravity}},  {\em JHEP} {\bf 07} (2010) 070,
  [\href{http://arxiv.org/abs/1005.0002}{{\tt arXiv:1005.0002}}].

\bibitem{Tanizaki:2019rbk}
Y.~Tanizaki and M.~\"Unsal, {\it {Modified instanton sum in QCD and
  higher-groups}},  {\em JHEP} {\bf 03} (2020) 123,
  [\href{http://arxiv.org/abs/1912.01033}{{\tt arXiv:1912.01033}}].

\bibitem{Pantev:2005rh}
T.~Pantev and E.~Sharpe, {\it {Notes on gauging noneffective group actions}},
  \href{http://arxiv.org/abs/hep-th/0502027}{{\tt hep-th/0502027}}.

\bibitem{Pantev:2005zs}
T.~Pantev and E.~Sharpe, {\it {GLSM's for Gerbes (and other toric stacks)}},
  {\em Adv. Theor. Math. Phys.} {\bf 10} (2006), no.~1 77--121,
  [\href{http://arxiv.org/abs/hep-th/0502053}{{\tt hep-th/0502053}}].

\bibitem{Pantev:2005wj}
T.~Pantev and E.~Sharpe, {\it {String compactifications on Calabi-Yau stacks}},
   {\em Nucl. Phys. B} {\bf 733} (2006) 233--296,
  [\href{http://arxiv.org/abs/hep-th/0502044}{{\tt hep-th/0502044}}].

\bibitem{Ohmori:2021fms}
K.~Ohmori, {\it {Replica Instantons from Axion-like Coupling}},
  \href{http://arxiv.org/abs/2101.07854}{{\tt arXiv:2101.07854}}.

\end{thebibliography}\endgroup

\end{document}